\documentclass{iaus}
\usepackage{graphics}

  \checkfont{eurm10}
  \iffontfound
    \IfFileExists{upmath.sty}
      {\typeout{^^JFound AMS Euler Roman fonts on the system,
                   using the 'upmath' package.^^J}%
       \usepackage{upmath}}
      {\typeout{^^JFound AMS Euler Roman fonts on the system, but you
                   dont seem to have the}%
       \typeout{'upmath' package installed. iaus.cls can take advantage
                 of these fonts,^^Jif you use 'upmath' package.^^J}%
      }
  \else
  \fi

  \checkfont{msam10}
  \iffontfound
    \IfFileExists{amssymb.sty}
      {\typeout{^^JFound AMS Symbol fonts on the system, using the
                'amssymb' package.^^J}%
       \usepackage{amssymb}%

      }{}
  \fi

  \IfFileExists{amsbsy.sty}
    {\typeout{^^JFound the 'amsbsy' package on the system, using it.^^J}%
     \usepackage{amsbsy}}
    {}

\newsavebox{\astrutbox}
\sbox{\astrutbox}{\rule[-5pt]{0pt}{20pt}}

\title[ --- short title of paper ---]
{Complex Topology of the Magnetic Field in Strong Flares}

\author[R.N.\,Ikhsanov, Yu.V.\,Marushin and N.R.\,Ikhsanov]
{R.N.\,Ikhsanov$^1$, Yu.V.\,Marushin$^1$ \and
N.R.\,Ikhsanov$^{1,2}$}

\affiliation{$^1$Pulkovo Observatory, St.\,Petersburg, Russia
email: solar@gao.spb.ru\\[\affilskip]
$^2$Korea Astronomy Observatory, Taejon, Republic of Korea email:
ikhsanov@kao.re.kr}

 \pubyear{2004}
 \volume{223}
 \pagerange{257-258}
 \date{22 and in revised form 33}
 \jname{Multi-Wavelength Investigations of Solar Activity}
 \editors{A.V. Stepanov, E.E. Benevolenskaya \& A.G. Kosovichev, eds.}

\begin{document}
\maketitle

\begin{figure}
 \centerline{\resizebox{12cm}{!}{\includegraphics{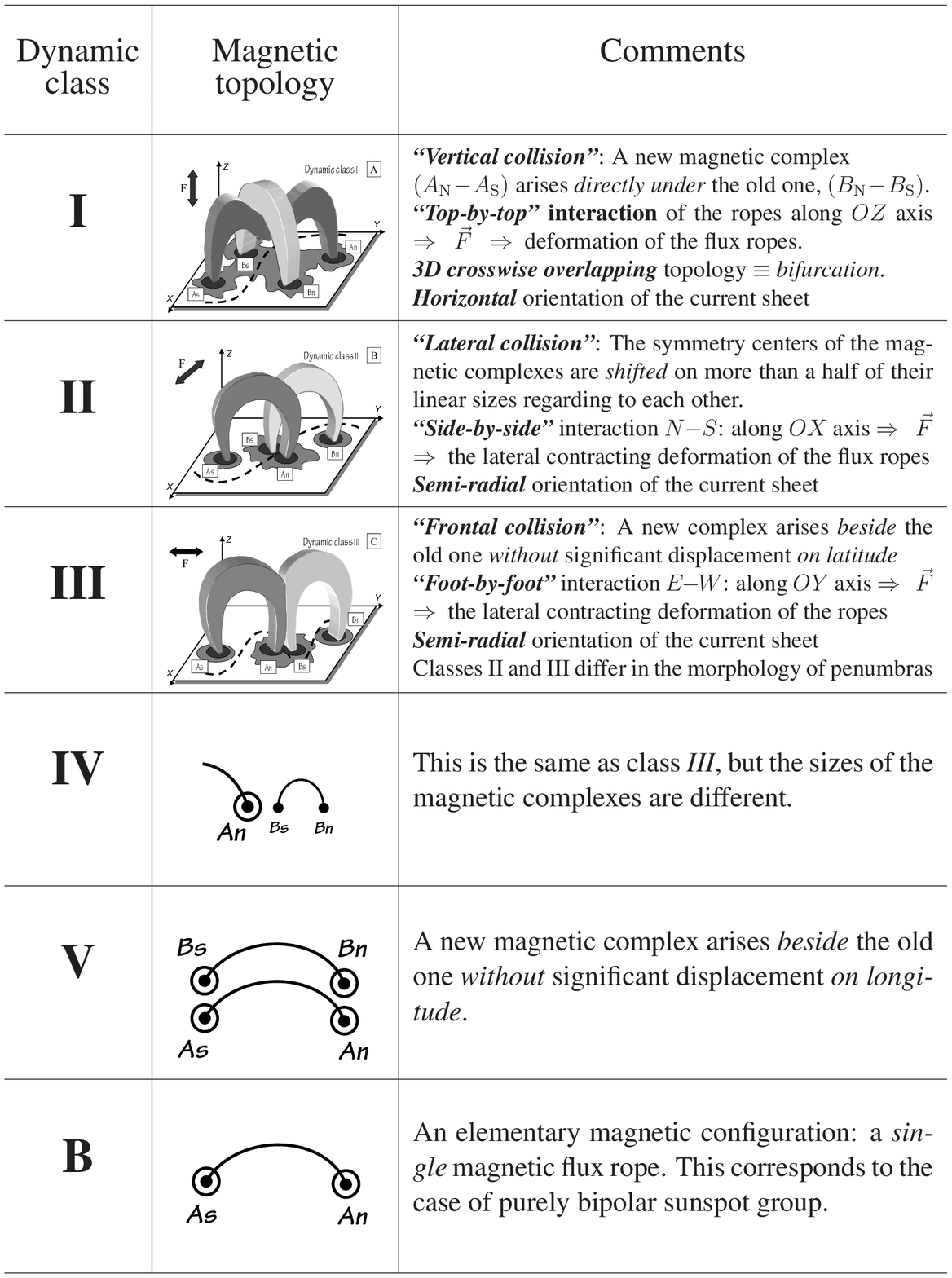}}}
  \caption{``5+1'' Dynamic classification of the field
    topology}\label{fig-1}
\end{figure}

We report the ``5+1'' dynamical classification of the most
frequently observed topologies of the magnetic field in sunspot
groups associated with powerful flares (see Fig.\ref{fig-1}). The
classification is based on the analysis of magnetographic and
H$\alpha$ observations of more than 600 active regions on the span
of 23\,years (for detailed description see \cite[Ikhsanov
1982]{Ikh82}, \cite[Ikhsanov 1985]{Ikh85}).

Statistical analysis of the processed data revealed 3\,basic types
of the magnetic field topology (dynamical classes I--III) in the
sunspot groups with $\delta$-configuration associated with
powerful (proton) flares (see Table\,2 in \cite[Ikhsanov \&
Marushin 2003]{IkhMar03} and references therein). Our results
confirm that the presence of $\delta$-configuration is a necessary
but not a sufficient condition for a strong flare in an active
region to occur. We find that all powerful flaring events (e.g.
proton flares) investigated within our project were related to the
emergence and development of the new magnetic flux tube(s) (or
complexes of tubes) in an active region, which strongly interact
with the magnetic complex earlier appeared in the region. It has
been also recognized (see e.g. \cite[Ikhsanov 1982]{Ikh82},
\cite[Ikhsanov 1985]{Ikh85}, and \cite[Ikhsanov \& Peregud
1988)]{IkhPer88} that the appearance of a flare depends on the
parameters of interacting flux tubes (i.e. the spatial scales and
the strength of the magnetic field) as well as on their mutual
position (e.g. the magnetic field topology). There is a strong
hint for a correlation between the flare parameters and the class
of the field topology realized in the active region to which the
flare is associated (\cite[Ikhsanov \& Peregud 1988]{IkhPer88}).
In particular, we find that the rate of distortions of the
interacting flux ropes is directly related to the rate of energy
release observed in corresponding flares.

Identification of the magnetic topology class  in an active region
provide us with information about possible location and spatial
orientation of current sheet in the flare associated with this
region. Application of this finding to the reconstruction of the
flaring scenario observed in the active region HR~17901+17906
(\cite[Ikhsanov \& Marushin 1998]{IkhMar98}) allows us to
 understand basically the observed anti-correlation between the
optical (H$\alpha$) and X-ray fluxes detected during two powerful
flares, which consequently appeared in this region. According to
our data, the value of the ratio (X-ray flux/optical flux) is
larger in the case of radially oriented current sheet, which
corresponds to the classes II and III of the magnetic topology
within our dynamical classification (see \cite[Ikhsanov \&
Marushin 2003]{IkhMar03} and references therein).

\begin{acknowledgments}
Nazar Ikhsanov acknowledges the support of Korea Astronomy
Observatory under their basic research programme. The work was
partly supported by the Russian Foundation of Basic Research under
the grant 03-02-17223a and the State Scientific and Technical
Program ``Astronomy''.
\end{acknowledgments}


\begin{thebibliography}{}

  \bibitem[Ikhsanov(1982)]{Ikh82}
  Ikhsanov, R.N. 1982, Izvestia GAO (Contributions of the Central Astronomical
  Observatory at Pulkovo), \textbf{200}, 15 (in Russian)

  \bibitem[Ikhsanov(1985)]{Ikh85}
  Ikhsanov, R.N. 1985, Izvestia GAO, \textbf{201}, 84 (in Russian)

  \bibitem[Ikhsanov \& Marushin(1998)]{IkhMar98} Ikhsanov, R.N.,
  Marushin, Yu.V. 1998, Izvestia GAO,  \textbf{212}, 91 (in Russian)

  \bibitem[Ikhsanov \& Marushin(2003)]{IkhMar03} Ikhsanov, R.N.,
  Marushin, Yu.V. 2003, astro-ph/0311114

  \bibitem[Ikhsanov \& Peregud(1988)]{IkhPer88} Ikhsanov, R.N.,
  Peregud, N.L. 1988, Solnechnye Dannye, \textbf{2}, 67 (in Russian)

\end{thebibliography}
\end{document}